\documentclass[conference]{IEEEtran}
\usepackage{graphicx}
\usepackage{tabularx}

\usepackage[T1]{fontenc}
\usepackage[utf8]{inputenc}
\usepackage[english]{babel}

\usepackage{csquotes}
\usepackage{nicefrac}
\usepackage{afterpage}
\usepackage{needspace}
\usepackage[textsize=scriptsize]{todonotes}
\usepackage{booktabs}
\usepackage{tikz}
\usepackage{quantikz}
\usetikzlibrary{arrows,positioning} 
\usetikzlibrary{shapes.geometric, positioning}
\usetikzlibrary{fit, quotes}
\usetikzlibrary{shapes.geometric, positioning}
\usetikzlibrary{automata,arrows}
\usepackage{flushend}
\usepackage{hyperref}
\usepackage{soul}
\usepackage{amssymb}
\usepackage{csvsimple}
\usepackage{amsthm}
\usepackage{lipsum}
\usepackage{varwidth}
\usepackage{utfsym}
\usepackage{fontawesome5}
\usepackage{multicol}
\usepackage{listings}
\usepackage{adjustbox}
\usepackage{float}
\usepackage{placeins}
\usepackage{yquant}
\usepackage{pgfplots}
\pgfplotsset{compat=1.18}
\usepackage[detect-all=true]{siunitx}
\usepackage{etoolbox}
\usepackage{bbm}
\usepackage{balance}
\usepackage{changes}
\usepackage{makecell}
\usepackage{microtype}

\robustify\bfseries

\makeatletter
\let\MYcaption\@makecaption
\makeatother
\usepackage[font=footnotesize]{subcaption}
\makeatletter
\let\@makecaption\MYcaption
\makeatother

\newtheorem{example}{Example}

\usepackage[backend=biber,
    style=ieee,
    citestyle=numeric-comp,
    sortlocale=en_US,
    sortcites=true,
    url=true,
    eprint=true,
    giveninits=false,
    dashed=false,
    minnames=1,
    maxnames=5]{biblatex}
\AtEveryBibitem{
  \clearname{editor}
  \clearfield{series}
  \clearfield{isbn}
  \clearfield{issn}
  \clearfield{volume}
  \clearfield{number}
  \clearfield{pages}
  \clearfield{doi}
}

\renewcommand{\baselinestretch}{0.975}

\hypersetup{
	pdftitle={Teaching Quantum Design Automation with Block-Based Programming},                      %
	pdfsubject={Quantum Science and Engineering Education Conference (QSEEC)},                    %
	pdfauthor={Damian Rovara, Robert Wille}    %
}

\newcommand{\block}[1]{``\emph{#1}''}
\newcommand{\link}{\url{https://munich-quantum-toolkit.github.io/scratch-quantum/}}
\newcommand{\github}{\url{https://github.com/munich-quantum-toolkit/scratch-quantum}}

\begin{document}

\renewcommand*{\figureautorefname}{Fig.}
\renewcommand*{\sectionautorefname}{Section}
\renewcommand*{\subsectionautorefname}{Section}
\def\exampleautorefname{Example}

\title{Teaching Quantum Design Automation\\with Block-Based Programming\vspace{-0.5em}}

\author{
\IEEEauthorblockN{Damian Rovara\IEEEauthorrefmark{1}\hspace*{2.4cm}Robert Wille\IEEEauthorrefmark{1}\IEEEauthorrefmark{2}}
\IEEEauthorblockA{\IEEEauthorrefmark{1}Technical University of Munich, Germany}
\IEEEauthorblockA{\IEEEauthorrefmark{2}MQSC, Garching near Munich, Germany}
\IEEEauthorblockA{\href{mailto:damian.rovara@tum.de}{damian.rovara@tum.de}\hspace*{1.5cm}\href{mailto:robert.wille@tum.de}{robert.wille@tum.de}\\
\url{https://www.cda.cit.tum.de/research/quantum}}
\vspace{-3.1em}
}

\maketitle

\begin{abstract}
As quantum circuits grow beyond small toy examples, preparing them for execution on physical devices becomes increasingly complex.
Design automation is therefore essential for scalable quantum computing:
\emph{Compilation procedures} optimize resource requirements and transform circuits to a format compatible with specific hardware; \emph{resource estimation} evaluates execution cost; \emph{verification methods} prove circuit correctness.
However, these concepts present a steep learning curve for novices, particularly when quantum circuits are introduced through low-level textual representations.
To address this, we present a block-based programming framework for quantum design automation, implemented as an extension to the \emph{Scratch} programming platform.
This system allows users to build quantum circuits as a sequence of blocks and embed them in classical control logic to perform evaluations, compare simulation results, and directly apply different design automation techniques.
We evaluated the approach in a user study with computer science students, who completed guided exercises using the platform and provided structured feedback in the form of self-reports and short knowledge assessments.
Results demonstrate strong understanding and confidence in quantum design automation concepts, suggesting that the block-based approach successfully lowers the entry barrier to quantum design automation.  
The implemented framework is open-source and available at~\mbox{\github}. 
\end{abstract}


\section{Introduction}
\label{sec:introduction}
Quantum computing is a rapidly growing field of research, built on a fundamental basis of interdisciplinarity.
Developers wanting to implement quantum algorithms require not only a strong understanding of the underlying quantum-mechanical principles.
They also need substantial knowledge in software development to understand the computational concepts that are necessary to build executable quantum programs.

However, at the same time, the steady improvements of real quantum hardware, such as the continuous reduction of gate errors and increases in qubit counts, also lead to more complex quantum programs.
This makes a fully manual development workflow error-prone and inefficient.
Preparing larger circuits for the execution on actual hardware, optimizing their resource requirements, or verifying their correctness present just some of many computationally hard tasks that are practically infeasible for manual processing.
When faced with the same problem in the classical world, circuit developers proposed \emph{design automation} methods that take over this manual workflow~\cite{wang2009}.

Unfortunately, the interdisciplinarity of quantum computing renders the adoption of these techniques difficult:
Many quantum computing experts do not have the necessary background in design automation, while many classical design automation researchers lack the required quantum-mechanical~fundament.

While several educational tools have been designed to ease the access for non-experts into the quantum computing world, there exists a distinct boundary between those tools and the potential applications in quantum design automation.
Graphical circuit editors, for instance, allow users to visually learn the effects of individual gates on underlying quantum states~\cite{composer, gayathridevi2023}.
However, these circuits are limited in their compatibility with other software tools and lack abstraction structures such as classical control flow, breaking the pedagogical link to actual software engineering.
Textual representations for quantum programs, such as quantum programming frameworks~\cite{koch2025, bergholm2022, qiskit, quirk} and intermediate representations (IRs, ~\cite{koch2025a, ittah2024, ittah2022, kirin, burgholzer2026a}), on the other hand, may address these limitation, at the cost of being more difficult to grasp for new learners.

In this work, we aim to bridge this gap by introducing a block-based framework for the development of quantum circuits as an extension to the \emph{Scratch}~\cite{resnick2009, maloney2010} platform.
This approach leverages the proven educational value of block-based development, while providing an easy integration of quantum and classical concepts through the pre-existing classical instructions provided by \emph{Scratch}.
Additionally, through a closely coupled translation to other IRs, we not only provide a stronger compatibility with the myriad of existing tools for quantum design automation, but also allow new learners to naturally mature into the larger quantum computing ecosystem.

An exploratory evaluation with computer science and electrical engineering students indicates that the tool lowers the entry barrier to quantum design automation concepts while maintaining high usability.
The proposed framework is fully open-source; its code is available at \github~as part of the \emph{Munich Quantum Toolkit}~(MQV,~\cite{willeMQTHandbookSummary2024}), with a web UI accessible at~\link.

The remainder of this work is structured as follows:
\autoref{sec:background} provides the required background in quantum design automation.
\autoref{sec:motivation} then presents in more detail the limitations of existing software tools to teach quantum design automation concepts and demonstrates how a block-based framework can address these limitations.
Then, \autoref{sec:a-quantum-extension-for-scratch} presents the core functionalities of the proposed framework, and \autoref{sec:exploring} explores how they can be used to teach quantum design automation concepts to new learners.
\autoref{sec:evaluation} then performs an empirical evaluation of the proposed framework and, finally, \autoref{sec:conclusion} concludes this work.

\section{Background}
\label{sec:background}

This section presents the fundamental concepts of quantum design automation and previous related work in the field.

\subsection{Design Automation in Quantum Computing}

As quantum algorithms grow in size, the manual preparation and evaluation of circuits is slowly becoming unfeasible even for experts in the field.
A similar trend in the classical circuit design community has previously led to the emergence of various electronic design automation methods, such as for circuit synthesis or verification~\cite{wang2009, molitor2010, brand1993, marques1999, jha1997}. 
As a result, design automation methods for quantum computing have also risen in popularity over the recent years.
In the following, we explore the core research areas of quantum design automation.

\paragraph*{Compilation}
Execution time on quantum hardware is expensive and longer circuit executions lead to a higher degree of noise, both due to qubit decoherence and gate errors.
Therefore, it is crucial to compile quantum circuits in an efficient way.
This is especially important when considering restrictive hardware properties, such as limited gate sets or qubit connectivities, that require input circuits to be transformed to fit the device characteristics.

Frameworks such as \emph{Qiskit}~\cite{qiskit} or \emph{TKET}~\cite{tket} are commonly employed software tools that aid in the compilation of quantum circuits.
They employ pattern rewrite rules to efficiently prune down circuits and prepare them for execution on quantum hardware.
At the same time, approaches based on machine learning~\cite{alexeev2025, quetschlich2025, fosel2021a, moro2021, tudisco2026} are also gaining popularity, taking advantage of massive datasets of quantum circuits that are available on different platforms~\cite{li2023, quetschlich2023} to train models to produce minimal representations of quantum circuits.

As new hardware types with different constraints---such as differing coupling maps or even entirely new qubit modalities---are being developed, compilation tools must be adapted and improved to tackle the relating challenges~\cite{li2019a, zilk2022, schmid2024}.
Furthermore, an increasing shift towards a hybrid ecosystem in which quantum and classical operations are required to be executed in an interleaved manner requires compilation toolchains that can handle these complex workflows.

\paragraph*{Resource Estimation}
Quantum hardware of the \emph{near-term intermediate scale quantum} (NISQ) era is limited both in accuracy and qubit count.
While smaller circuits can be executed on this hardware, larger problem instances quickly reach the limits of current devices.
Similarly, state-of-the-art simulators~\cite{kang2025, bayraktar2023, javadi-abhari2024, grurl2022, zulehner2018} also struggle with the execution of arbitrary quantum circuits of larger sizes.
As a result, reasoning over large circuits becomes challenging.

To still be able to evaluate large quantum circuits, \emph{resource estimation} tools are commonly employed.
They analyze circuits in terms of their resource requirements and yield results that can be used to determine the performance of optimizations or compare hardware vendor roadmaps to estimate at what time we can expect advanced quantum computers to execute these circuits~\cite{forster2025a, forster2025b}.
Resource estimation tools often also investigate further metrics, such as accuracy estimates~\cite{suchara2013} or approximate costs for the fault-tolerant execution of the circuit~\cite{vandam2023using}.

\paragraph*{Verification}
Methods for verification play an important role in evaluating the correctness of complex quantum circuits.
Quantum Hoare logic~\cite{zhou2019}, for instance, provides a fundamental framework for the formal verification of circuits, extending classical verification rules to the quantum domain.
Assertions are furthermore being employed for the verification and testing of quantum circuits~\cite{huang2019, li2020, rovara2025a, ying2022, liu2020, liu2021, witharana2023, rovara2025b, rovara2025c}, allowing circuit designers to specify the expected quantum state during the execution and comparing it with execution outcomes.

Finally, quantum circuit equivalence checking has been commonly employed, especially to verify the correctness of circuit transformations~\cite{yamashita2010, burgholzer2021a}.
As empirical evaluations have shown the prevalence of bugs in quantum circuits, verification and testing is becoming an increasingly important aspect of quantum design automation~\cite{upadhyay2025a, dimatteo2024, metwalli2024, miranskyy2021, garciadelabarrera2023}.

\medskip

As the design and evaluation of quantum circuits is a conceptually difficult topic, educational tools require visual aids to provide an understanding of the underlying concepts.
The following evaluates related software tools in the field, investigating their use cases.

\subsection{Software Tools for Quantum Design Automation}

With the growing need for design automation tools in quantum computing, several larger software tools have found wide-spread adoption.
The \emph{Qiskit} framework~\cite{qiskit}, for instance, can be used to construct quantum circuits, to optimize and analyse them, as well as to execute them through simulators or on physical devices.
Similarly, \emph{TKET}~\cite{tket} is another popular software tool for the compilation of quantum circuits.
The \emph{Munich Quantum Toolkit}~\mbox{(MQT,~\cite{willeMQTHandbookSummary2024})} is another set of design automation software tools for quantum circuits, ranging from high-level application scenarios~\cite{quetschlich2024, rovara2026a} to device-specific and device-agnostic compilation passes~\cite{wille2023, rovara2026, rovara2025d}.

The \emph{Microsoft Azure Quantum Development Kit}~\cite{prateek2023} provides a programming environment for the design of quantum circuits, as well as several tools for debugging and testing.
Through its quantum resource estimator~\cite{bergholm2022, vandam2023using}, it also provides users with the ability to evaluate the resource requirements for the fault-tolerant execution of quantum circuits.

All of these tools are compatible with a variety of low-level quantum programming languages, such as OpenQASM~\cite{cross2022}.
At the same time, other compilation frameworks are instead building on classical compilation infrastructure, predominantly based on or inspired by the \emph{Multi-Level Intermediate Representation} (MLIR,~\cite{lattner2021}) framework~\cite{burgholzer2026a, ittah2024, koch2025a, kirin}.
Further higher-level quantum development kits, such as \emph{PennyLane}~\cite{bergholm2022}, \emph{Qrisp}~\cite{seidel2024}, or \emph{Guppy}~\cite{koch2025}, build on top of these quantum-classical representations.

\section{Motivation}
\label{sec:motivation}

This section highlights core challenges that emerge when teaching quantum design automation concepts.
It then proposes a new approach to more easily and intuitively teach these concepts to students.

\subsection{Challenges in Teaching Quantum Computing}

\begin{figure*}
    \centering
    \includegraphics[width=\textwidth]{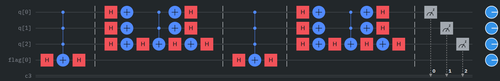}
    \caption{Implementing Grover's algorithm using IBM's \emph{Quantum Composer} after unrolling the loop into a single circuit.}
    \label{fig:grover-composer}
\end{figure*}

Textual code representations for quantum circuits, such as \emph{OpenQASM}~\cite{cross2022}, \emph{QIR}~\cite{qir}, or \emph{MLIR}~\cite{lattner2021} find a lot of applications in quantum design \mbox{automation~\cite{burgholzer2026a, hopf2026}}. %
However, reading the resulting code files and gaining an intuitive understanding of their inner workings poses a challenge even for experts with years of experience in the field.
Asking new learners to start with these verbose and complex representations would greatly hinder their process, which is why visual quantum circuit editors have risen in popularity over the last years.
Tools such as \emph{Quirk}~\cite{quirk, gallardo2024} or IBM's \emph{Quantum Composer}~\cite{composer, gayathridevi2023} allow learners to easily get acquainted with quantum circuits by providing a drag and drop interface in which users can add gates to a circuit and, through integrated simulation methods, view how changes to the circuit affect the resulting state.

However, as quantum computing research has grown into a much broader interdisciplinary field---incorporating ideas such as HPC integration or fault tolerance---circuit editors have started to demonstrate several shortcomings that limit the degree to which they can cover the quantum computing~ecosystem.

Notably, the introduction of \emph{quantum-classical interactions} in quantum computing workflows slowly leads to a deviation from typical circuits. %
Dynamic measurements are starting to be more commonly included in the standard repertoire of supported operations~\cite{corcoles2021, pino2021, egger2018, mcclure2016}.
Even further, other control flow features, such as \emph{loops} and \emph{function calls}, are completely incompatible with the current visual circuit representation methods.

Furthermore, while visual circuit editors might be a useful pedagogical stepping stone for circuit design methods, their integration with other design automation tools is often very loose.
IBM's \emph{Quantum Composer}, for instance, allows the export to and import from \emph{OpenQASM} programs.
However, to run and evaluate transformations on the generated circuits, users are required to move away from the visual editor.
This boundary between circuit design and other design automation tasks renders the topics very remote and hinders the natural integration of the ideas behind them.

Finally, as the landscape of automated tools in the field of quantum computing grows wider, so does the required compatibility of generated quantum programs.
While \emph{OpenQASM} is commonly employed, many software frameworks also build upon \emph{QIR}, \emph{MLIR}, or other intermediate \mbox{representations~\cite{burgholzer2026a, koch2025, koch2025a, ittah2024, ittah2022, kirin}}.
To accelerate the education process of new learners and maximize the tools they have at their disposal, quantum program design tools require compatibility with as many different representations as possible.

\begin{example}
\label{ex:1}
Consider teaching Grover's algorithm~\cite{grover} via a standard visual circuit editor. 
As the algorithm relies on the repeated application of \emph{oracle} and \emph{diffusion} steps, the use of loops would provide an intuitive understanding of this idea to students with a computer science background.
However, because existing tools typically lack classical control structures, the algorithm must be manually unrolled, stripping away a crucial layer of algorithmic abstraction.
This is illustrated in~\autoref{fig:grover-composer} for just two loop iterations.
Most of the visualization is taken up by two identical circuit segments.
An implementation with more iterations will lead to even larger redundancies in the visualization, substantially harming the readability.
Furthermore, demonstrating the algorithm's flexibility for different search targets requires the student to manually rearrange the underlying circuit for every new search value.
Lastly, if we then want to analyze the resource costs of these different variations, or investigate potential optimizations to the circuit, it must be repeatedly exported to external tools (e.g., via OpenQASM). 
This creates a disjointed workflow severely hindered by constant, repetitive manual export cycles.
\end{example}

This example shows that, to fully facilitate the learning process of developers interested in quantum computing, we need to provide a wider playground that makes development easy while providing the maximum degree of compatibility with design automation methods. 

\subsection{Proposed Framework}

To address the issues and limitations illustrated above, we propose adopting a proven methodology from classical computer science education.
Specifically, we introduce a framework that extends \emph{Scratch}---a widely used block-based programming platform that provides a drag-and-drop environment in which users can build programs by connecting individual instruction blocks~\cite{resnick2009, maloney2010}.
While preliminary efforts have previously explored introducing quantum concepts into Scratch~\cite{escanez2025, rosa2024}, these approaches largely focus on higher-level quantum mechanical concepts, and lack the underlying architectural infrastructure required for quantum design automation.
As an extension to \emph{Scratch}, the proposed framework can take advantage of a wide variety of classical programming concepts, provided in a beginner-friendly format to give new students easier access~\cite{armoni2015}.
In particular, this also includes classical control flow such as loops, conditionals, and function calls.

The proposed extension provides a set of quantum computing instructions, such as for the allocation of qubits and application of gates and measurements.
This allows users to construct quantum programs the same way they would using a visual circuit editor while also having full access to classical~computation.

Furthermore, the extension allows the simulation of the constructed circuits with a modular simulation backend, which enables easy extensibility with different simulator types.
In addition, further analysis tools, such as different \emph{resource estimation} methods, are directly integrated in the framework, so that no interaction with external tools is strictly required.
On the other hand, if interoperability with other tools and frameworks is desired, the proposed framework also supports the export of constructed circuits into an MLIR-based format which allows the translation to commonly used formats such as \emph{OpenQASM} and \emph{QIR}, as well as to the \texttt{jeff} exchange format~\cite{jeff}, through which compatibility with a variety of quantum program compilation and execution pipelines is enabled~\cite{burgholzer2026a, koch2025, koch2025a, ittah2024, kirin}.
This allows learners to naturally grow into the mature quantum computing~ecosystem.

\begin{figure}
    \centering
    \includegraphics[width=\columnwidth]{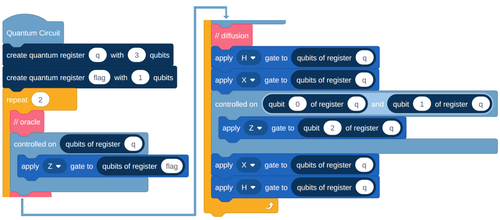}
    \caption{Implementing Grover's algorithm inside the proposed \emph{Scratch} extension.}
    \label{fig:grover-scratch}
\end{figure}

\begin{example}
Revisiting the Grover's algorithm scenario from \autoref{ex:1}, the proposed block-based framework eliminates the issues.
As \emph{Scratch} supports classical control flow structures, the loop that had to be unrolled for circuit editors can instead be applied intuitively.
An example implementation of the circuit is illustrated in~\autoref{fig:grover-scratch}.
As can clearly be seen, the algorithm itself is only defined once, while the iterations themselves are covered by the \block{repeat} block.
Similarly, classical instructions and conditionals can be used to execute the algorithm for various inputs, allowing users to search for different values with only minimal changes to the program.
Finally, as resource estimation is directly integrated into the system, no export to other frameworks is required.
Resource estimation is simply performed through the switch of a single instruction, changing between ``\emph{simulation}'' and ``\emph{estimation}'' mode.
This way, even the evaluation of different versions of the program can always be performed with just a single press of the ``\emph{play}'' button.
\end{example}

\section{A Quantum Extension for Scratch}
\label{sec:a-quantum-extension-for-scratch}

This section describes the technical details of the proposed framework, including the usage of the different features it supports, as well as how they are implemented.
The implementation of the framework is open-source and a live version is accessible at \link.

\subsection{Quantum Circuits in the Scratch Extension}

Many of the issues listed above, such as the limited classical methods available in circuit simulators, are already addressed by the strong foundation provided by the \emph{Scratch} framework.
It provides instructions to create and modify variables, control the execution flow, or even interact with the user.
Therefore, the proposed extension only needs to implement instructions for the actual creation of quantum circuits.
This includes:\begin{itemize}
    \item An instruction to \emph{allocate named registers} of qubits with arbitrary sizes.
    \item Instructions to \emph{apply different gates} to a provided circuit (including parametrized rotational gates).
    \item \emph{Control modifier blocks} that can take on child instructions that will be controlled by the given qubit, similarly to how other higher-level languages implement control modifiers~\cite{cross2022, seidel2024, burgholzer2026a}. Using further \block{negated control} and \block{multi-control} blocks also allows more complex controlled gates.
    \item A \block{measurement} block that returns a classical value which can be easily integrated into the classical execution environment of the standard \emph{Scratch} instructions.
\end{itemize}

Quantum operations can only be placed below a \block{Quantum Circuit} parent block.
This allows users to implement several independent quantum circuits and compare them more easily.
All quantum operations allow variable inputs to be specified with dynamic attributes, such as variables or computation results.
This allows a simple and direct interaction between classical computations (e.g., to decide target qubits for operations) and the quantum circuit.

\begin{figure}
    \centering
    \includegraphics[width=0.8\columnwidth]{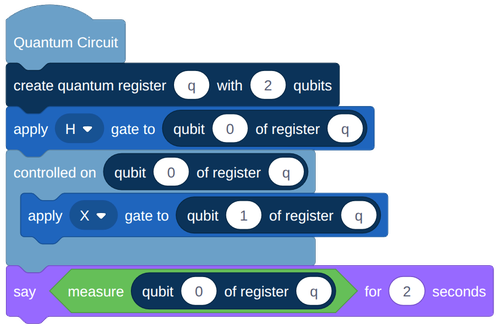}
    \caption{A Bell circuit created through the \emph{Scratch} extension.} \vspace{-1em}
    \label{fig:circuit}
\end{figure}
\begin{example}
    \autoref{fig:circuit} illustrates the preparation of a Bell state using the proposed framework.
    It starts with the allocation of a qubit register consisting of two qubits.
    It then applies a Hadamard gate to the qubit at index 0 ($q_0$), followed by a \texttt{CX} gate controlled on $q_0$.
    Finally, it performs a measurement of $q_0$ and passes the output to the standard \emph{Scratch} block \block{say [text] for [time] seconds} to display the measurement result.
\end{example}

With these quantum computing concepts integrated into the \emph{Scratch} development platform, the next important step is to provide support for quantum design automation methods.
This is presented next, more precisely through simulation, resource estimation, and further compatibility with more advanced methods. 

\subsection{Modular Simulator Backend}

The proposed framework allows for the modular application and exchange of different simulation backends to execute the provided circuits.
To this end, it defines an interface specifying a set of functions that have to be implemented by a compatible simulator, such as register allocation, gate application, and state measurement.
During the execution of the \emph{Scratch} program, the quantum-specific blocks will only access these interface methods of the simulator.
By relying on this abstraction, the architecture natively supports the runtime exchange of simulation backends.
While the current implementation features only a single state vector simulator, this modular design provides a foundation for future extensions where learners could seamlessly investigate and compare the performance of different simulation methodologies.

To start the simulation process, the \block{Run Quantum Circuit} block can be used.
This instruction causes all circuits contained in the current program to be executed in parallel.
Simulations can be performed repeatedly, for instance to emulate the evaluation of several shots of a circuit.
Furthermore, the \block{Run Quantum Circuit starting from [state]} block can also be used as an alternative.
This instruction starts all simulations, but initializes the quantum state with the given value.
This allows users to modularly execute quantum circuits from different starting states. 

\begin{figure}
    \centering
    \begin{subfigure}{0.45\columnwidth}
        \centering
        \includegraphics[width=\textwidth]{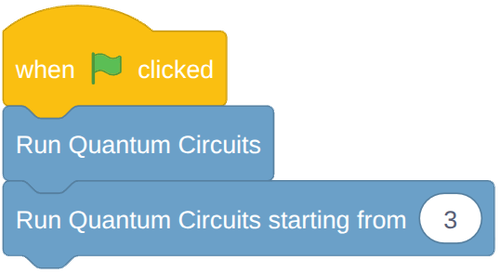}
        \caption{}
        \label{fig:simulation}
    \end{subfigure}
    \hfill
    \begin{subfigure}{0.45\columnwidth}
        \centering
        \includegraphics[width=0.8\textwidth]{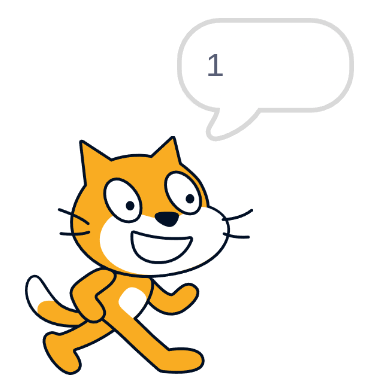}
        \caption{}
        \label{fig:simulation-out}
    \end{subfigure}
    \caption{Starting the simulation of the circuit in~\autoref{fig:circuit} from the \emph{Scratch} framework and the output of the final simulation.}
\end{figure}

\begin{example}
\autoref{fig:simulation} shows the main entry point of a \emph{Scratch} program, prepared to execute the circuit defined in~\autoref{fig:circuit}.
The circuit is first executed starting implicitly from the $\ket{00}$~state.
In a second step, the circuit is instead executed starting from the state $3$, which is equivalent to the $\ket{11}$ state.
The corresponding output, which is displayed using the \block{say [text] for [time] seconds} printing interface, is shown in~\autoref{fig:simulation-out}.
In both cases, the measured qubit is located with equal probability in both computational basis states.
Accordingly, executing the program multiple times, yields both potential results equally as often.
\end{example}

At the moment, the default simulator implemented for the proposed framework allows the allocation of up to 10 qubits.
Attempting to allocate further qubits safely halts execution to prevent memory exhaustion on the host machine.
Analyzing larger algorithms therefore requires alternative methods, which we address through the framework's integrated \emph{resource estimation}.
\subsection{Resource Estimation}

By substituting the \block{Run Quantum Circuit} block with a \block{Perform Quantum Resource Estimation} block, provided circuits are instead analyzed for their resource requirements.
As the provided circuits may depend on dynamic runtime values and can use control flow instructions such as dynamic loops, the required resources cannot be analyzed statically.
Instead, the proposed framework employs a stochastic tracing \mbox{method~\cite{ittah2022, meuli2020}}. %
Rather than tracking a full quantum state vector, it only stores an approximate state of individual qubits.
Whenever a measurement is encountered, an outcome is chosen randomly based on the information tracked by this method.
While this does not always lead to correct measurement outcomes, it generates an approximate execution trace that is often indicative for the normal execution of the provided circuit.

This trace is then used to compute the required number of resources, such as qubit count or gate counts.
The estimated resources are then stored in classical variables that can be accessed outside of the circuit to investigate analysis results.
This allows learners to immediately access resource estimation results without requiring them to move to a different framework.
Furthermore, they can just as easily compare the impact of changes to circuits by modifying them and immediately rerunning the estimation.
As the stochastic tracing method does not suffer from the same exponential runtime explosion required for full simulation, this method can be used to reason over larger programs that cannot be simulated.

To provide an overview of different potential cost models for quantum circuits to learners, the proposed framework also allows different \emph{decomposition settings} to be specified.
For instance, by enabling \emph{\texttt{SWAP} decomposition}, \texttt{SWAP} gates are internally replaced by three \texttt{CX} gates, as is often required for the compilation of quantum circuits for physical devices.
This allows learners to more easily visualize the increased cost of more expensive gates.
Furthermore, \emph{Toffoli decomposition} can be enabled to force all multi-controlled gates to instead be decomposed into several single and two-qubit gates.
This decomposition employs additional ancilla qubits to reduce the number of operations required for decomposition~\cite{nielsen2010}.
This way, learners can experience not just an increase in gate count but also qubit count as they implement circuits with multi-controlled gates.
Overall, these decomposition settings allow users to explore the bottlenecks of larger quantum circuits, as well as potential targets for circuit optimizations.

\begin{figure}
    \centering
    \includegraphics[width=0.44\columnwidth]{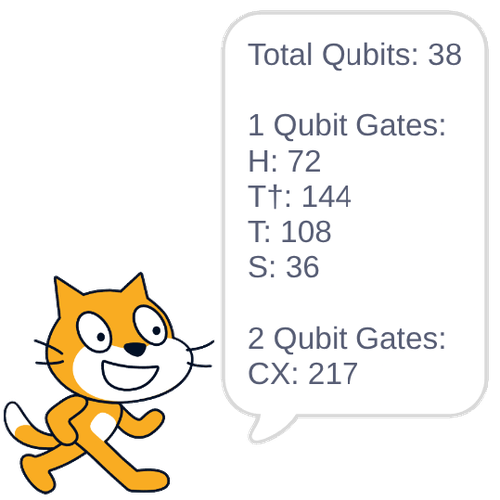}
    \caption{The resource estimates for the \emph{Toffoli decomposition} of a 20-qubit Toffoli gate.}
    \label{fig:re}
\end{figure}

\begin{example}
Consider a quantum circuit using 20 qubits that applies a Hadamard gate to 19 of these qubits using a simple loop, and then performs a Toffoli gate using these qubits as controls and targeting the final qubit.
While this circuit is too large for full simulation, resource estimation correctly determines that this circuit requires 20 total qubits and uses 19 single-qubit Hadamard gates and one Toffoli gate on 20 qubits, correctly counting the resources even in the presence of a loop structure.
If we instead enable Toffoli decomposition and perform the resource estimation again, the estimation yields 38 total qubits with 379 single-qubit gates and 217 two-qubit gates.
The exact output resulting from displaying the resource estimates are shown in~\autoref{fig:re}.
This outcome appropriately highlights the resource costs of Toffoli gates with several control qubits.
\end{example}

\subsection{Compatibility with other Methods}

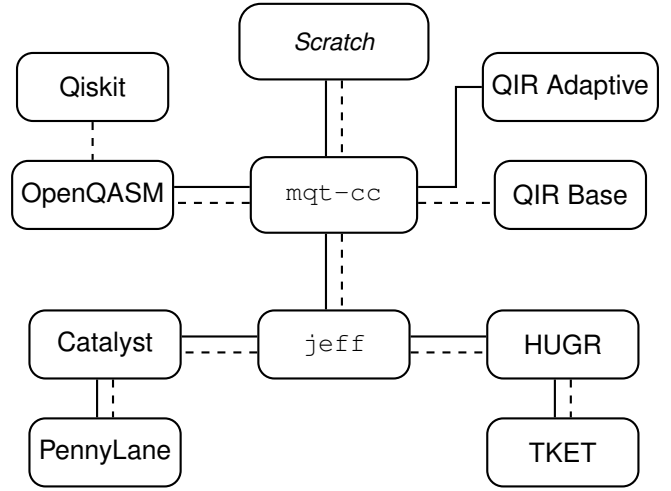
\begin{figure}
    \centering
    \begin{tikzpicture}[
  node distance=1.5cm and 2.2cm,
  every node/.style={font=\sffamily},
  mainbox/.style={draw, rounded corners=6pt, minimum width=2cm, minimum height=0.9cm, align=center, thick},
  dashbox/.style={draw, dashed, rounded corners=6pt, minimum width=2cm, minimum height=0.9cm, align=center, thick},
  cloudbox/.style={draw, ellipse, minimum width=2cm, minimum height=0.9cm, align=center, thick},
  dashcloud/.style={draw, dashed, ellipse, minimum width=2cm, minimum height=0.9cm, align=center, thick},
  solidline/.style={-,thick},
  dashline/.style={-,dashed,thick},
]

\node[draw, rounded corners=8pt, thick, minimum width=2.5cm, minimum height=1.0cm] (topgroup) at (3,7.5) {};
\node[font=\small\sffamily] at (3,7.5) {\emph{Scratch}};

\node[mainbox, minimum width=2.2cm, minimum height=1.0cm, below=1cm of topgroup] (mqtcc) {\texttt{mqt-cc}};

\node[mainbox, left=1.0cm of mqtcc] (qasm) {OpenQASM};
\node[mainbox, above=0.5cm of qasm] (qiskit) {Qiskit};

\node[mainbox, right=1.0cm of mqtcc] (qirbase) {QIR Base};
\node[mainbox, above=0.5cm of qirbase] (qiradaptive) {QIR Adaptive};

\node[mainbox, below=1.0cm of mqtcc] (jeff) {\texttt{jeff}};

\node[mainbox, left=1.0cm of jeff] (catalyst) {Catalyst};
\node[mainbox, right=1.0cm of jeff] (hugr) {HUGR};
\node[mainbox, below=0.5cm of hugr] (tket) {TKET};
\node[mainbox, below=0.5cm of catalyst] (pennylane) {PennyLane};

\draw[solidline] ([xshift=-3pt]topgroup.south) -- ([xshift=-3pt]mqtcc.north);
\draw[dashline] ([xshift=3pt]topgroup.south) -- ([xshift=3pt]mqtcc.north);

\draw[dashline] ([yshift=-3pt]mqtcc.west) -- ([yshift=-3pt]qasm.east);
\draw[solidline] ([yshift=3pt]mqtcc.west) -- ([yshift=3pt]qasm.east);

\draw[dashline] (qiskit.south) -- (qasm.north);

\draw[dashline] ([yshift=-3pt]mqtcc.east) -- ([yshift=-3pt]qirbase.west);

\draw[solidline] ([yshift=3pt]mqtcc.east) -- ++(0.5,0) |- (qiradaptive.west);

\draw[solidline] ([xshift=-3pt]mqtcc.south) -- ([xshift=-3pt]jeff.north);
\draw[dashline] ([xshift=3pt]mqtcc.south) -- ([xshift=3pt]jeff.north);

\draw[dashline] ([yshift=-3pt]jeff.west) -- ([yshift=-3pt]catalyst.east);
\draw[solidline] ([yshift=3pt]jeff.west) -- ([yshift=3pt]catalyst.east);

\draw[dashline] ([yshift=-3pt]jeff.east) -- ([yshift=-3pt]hugr.west);
\draw[solidline] ([yshift=3pt]jeff.east) -- ([yshift=3pt]hugr.west);

\draw[solidline] ([xshift=-3pt]hugr.south) -- ([xshift=-3pt]tket.north);
\draw[dashline] ([xshift=3pt]hugr.south) -- ([xshift=3pt]tket.north);

\draw[solidline] ([xshift=-3pt]catalyst.south) -- ([xshift=-3pt]pennylane.north);
\draw[dashline] ([xshift=3pt]catalyst.south) -- ([xshift=3pt]pennylane.north);

\end{tikzpicture}
    \caption{Interactivity of \emph{Scratch}-\emph{quantum} programs with other frameworks. 
    Dashed lines indicate the translation of purely quantum programs while solid lines also include quantum-classical programs.}
    \label{fig:compatibility}
    \vspace{-1em}
\end{figure}

To simplify interoperability with other frameworks, the block \block{Export Circuits to MLIR} can be used to translate a given circuit into an MLIR-based format provided by the \emph{mqt-cc} compilation framework~\cite{burgholzer2026a}.
This framework provides translation methods to many other commonly employed quantum computing tools.
\autoref{fig:compatibility} illustrates the different representations that quantum circuits constructed in the proposed framework are compatible with.

Notably, circuits that have been implemented using only quantum instructions can easily be translated to \emph{Qiskit}~\cite{qiskit} \texttt{QuantumCircuit} objects.
This allows the integration with the mature python-based quantum circuit framework.
More complex programs, particularly if they involve control flow structures, can instead be translated to the adaptive QIR profile~\cite{qir}.
Furthermore, by going through the \texttt{jeff} exchange format~\cite{jeff}, they can also be used within various other compilation frameworks, such as \emph{Catalyst}~\cite{ittah2024} or \emph{TKET}~\cite{tket}.
Through newer models of quantum hardware, such as the \emph{Helios} quantum computer~\cite{ransford2025}, such programs with dynamic structured control flow can even be executed on physical quantum hardware.
This wide range of interoperability is core to providing learners with a launching pad to get more closely involved with the quantum computing ecosystem.

\section{Exploring Quantum Design Automation\\with Scratch}
\label{sec:exploring}

Having established the proposed framework's technical architecture, this section demonstrates its pedagogical utility.
To illustrate how the extension facilitates the understanding of quantum design automation, we present a guided workflow centered around a single running example illustrated in~\autoref{fig:running-example-circuit}.
This workflow mirrors a realistic engineering lifecycle: moving from initial circuit design and resource estimation to optimization and verification.
By stepping through this process, we highlight how the integrated environment allows learners to directly observe the hardware-level consequences of their design choices without relying on disconnected external tools.

\subsection{Circuit Design}

A key advantage of using the proposed framework is its integration with classical computation concepts.
This allows learners to not just build quantum circuits, but instead implement classical programs that \emph{contain} quantum circuits.
To easily distinguish between the types of instructions, \emph{Scratch} employs a color-coding system.
By assigning different shades of blue colors to quantum instructions, we can create a strong visual boundary to classical control flow and variable storage instructions, which instead largely use orange and yellow tones.
This enables learners to translate standard circuit diagrams or textual algorithm descriptions into Scratch programs intuitively, easily distinguishing quantum operations from classical logic at a glance.

 \begin{figure}
    \centering
    \includegraphics[width=0.4\textwidth]{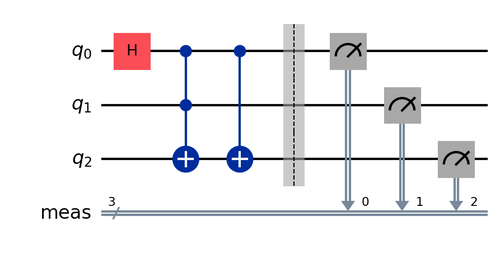}
    \vspace{-0.5em}
    \caption{A simple quantum circuit with three qubits, used as a running example for this section}
    \label{fig:running-example-circuit}
    \vspace{-0.5em}
 \end{figure}

\begin{figure*}
 \centering
 \hspace{-1.3cm}
 \begin{subfigure}{0.28\textwidth}
    \centering
    \includegraphics[width=\textwidth]{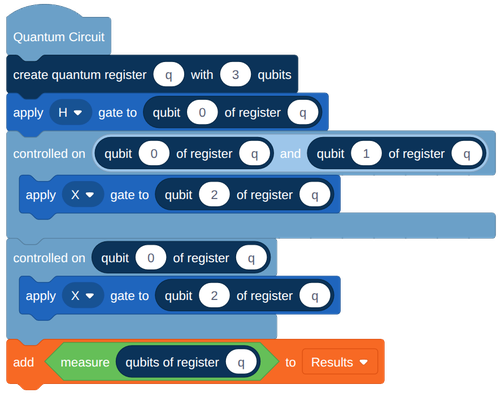}\\\vspace{0.1em}
    \phantom{hello}\\
    \caption{Initial circuit implementation.}
    \label{fig:running-example-scratch}
 \end{subfigure} \phantom{hi}
 \begin{subfigure}{0.65\textwidth}
    \centering
    \includegraphics[width=1.1\textwidth]{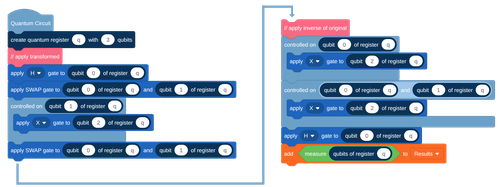}
    \caption{Equivalence checking.}
    \label{fig:verification}
\end{subfigure}
 \caption{A quantum circuit with three qubits, implemented and verified in \emph{Scratch}. 
 (a) Implementing the baseline circuit defined in~\autoref{fig:running-example-circuit}. 
 All qubits are measured and the outcomes are stored in the classical list Results. 
 (b) Checking the equivalence of the transformed circuit by concatenating it with the inverse of the original. 
 If the Results list contains at least one entry with the value 1, the circuits are not equivalent for the starting state |000>.}
 \label{fig:running-example}
\end{figure*}

\begin{example}
\autoref{fig:running-example-circuit} illustrates a quantum circuit using 3 qubits.
Students that have already been introduced to the concept of quantum circuits can easily be asked to translate this circuit into a \emph{Scratch} program by dragging in the corresponding blocks.
The resulting \emph{Scratch} program is shown in~\autoref{fig:running-example-scratch}.
This example illustrates concepts such as controlled gates and measurements.
The shift from the ``quantum world'' to the ``classical world'' through measurements, with outcomes stored in a classical list of bits, is clearly visualized with the change in the hue of the blocks, ranging from the blue quantum-related blocks to the orange blocks for data storage, bridged together by the green measurement bloks.
Learners can now be asked to simulate the circuit and view the measurement outcomes directly through the built-in variable inspection system or to work with the measurement outcomes in further classical instructions.
\end{example}

\subsection{Evaluation with Resource Estimation}

After constructing the quantum circuit, we next pose the question of how its performance can be evaluated.
By giving learners the freedom to modify the circuit, including the addition of further qubits, they may already experience a decrease in responsiveness as circuits grow larger.
However, once the limits of the underlying simulator are crossed, this indirect evaluation process will no longer work.
At this point, we introduce the concept of resource estimation to address the limitations of simulation-based analysis:\begin{itemize}
    \item As simulating the circuit is not required, it can efficiently process circuits of various sizes with hundreds of qubits.
    \item It focuses on specific metrics that are more meaningful for actual execution that is impacted by differences in errors and execution times of different gates.
    \item It can specifically address hardware constraints such as limited gate compatibilities. 
\end{itemize}

Resource estimation allows learners to investigate the metrics most relevant to physical hardware constraints.
Furthermore, by covering different gate decomposition types, learners can intuitively grasp the cost differences individual gates carry.

\begin{example}
    Returning to the example program shown in \autoref{fig:running-example}, we ask students to perform resource estimation by replacing the \block{Run simulation} block with a \block{Perform Quantum Resource Estimation} instruction.
    This process reports a total qubit count and gate count of 3.
    After enabling \emph{Toffoli decomposition} for this program, the number of required qubits rises to 4 and the number of gates rises to 35.
    In particular, this includes a total of 14 $T$ and $T^{-1}$ gates, as well as 14 $CX$ gates.
    Finally, we ask students to add additional control qubits to the Toffoli gate and investigate the changes in resource requirements.
    This way, learners can independently discover the relation between control counts in Toffoli gates and total qubit, as well as gate counts.
\end{example}

\subsection{Compilation}

While certain quantum circuit compilation tasks, such as mapping operations to a native gate level, were already covered in the previous step on resource estimation, \emph{circuit optimization} still represents an important aspect of compilation.
After being given performance evaluation methods in the form of resource metrics, learners can now attempt to improve those metrics by applying changes to the circuit.
Furthermore, the concept of qubit routing can be introduced at this stage, using the \block{apply \texttt{SWAP} gate} instruction and asking learners to prepare circuits for a given coupling map.
By then reevaluating the resource requirements, students can determine how their proposed optimizations, as well as the overhead introduced by inserting the \texttt{SWAP} gates, impacted the circuit's performance~metrics.

\begin{example}
\label{ex:compilation}
We highlight how the $q_1$ in \autoref{fig:running-example} will stay in the $\ket{0}$ state throughout the entire execution of the circuit.
As a result, the first Toffoli gate in the circuit effectively has no effect on the overall quantum state, as $q_1$ is one of its control qubits, and can therefore safely be removed.
Further, we introduce a coupling map with nearest neighbor connectivity among individual qubits.
As $q_0$ and $q_2$ are part of a shared $CX$ gate, the current implementation of the circuit violates the connectivity constraints.
By introducing a \texttt{SWAP} gate between $q_0$ and $q_1$ before the $CX$ gate, as well as a second \texttt{SWAP} afterwards to return to the starting configuration, learners can compile the sample program for the simple coupling map.  
\end{example}

\vspace{-0.5em}
\subsection{Verification}
\vspace{-0.5em}

An important procedure that is part of the quantum design automation pipeline is the verification of circuits.
It is required to show that changes made to the circuit during the compilation process did not introduce errors.
One popular approach for the verification of compilation outcomes is by performing an equivalence check with the previous state of the circuits~\cite{yamashita2010, burgholzer2021a}.
For unitary quantum circuits, this can be achieved by concatenating the inverse of the original circuit to the application of the updated circuit.
If this reduces the state to its starting state, the circuits are equivalent and the transformations were applied correctly.

In the context of the proposed framework, performing this check is particularly easy:
If the considered circuit only uses self-inverse operations, users can simply drag the operation blocks from the starting circuit to the end of the new circuit in reverse order.
By then measuring all qubits and displaying the outcomes over several repeated simulations, students can interactively experience this verification process and gain an intuitive understanding of its workings.
In cases where an introduced error only amounts to a probabilistic difference in the final measurement outcomes, they experience how the accuracy of this method depends on the number of performed simulations.
By performing the simulations from different starting states, learners can also see how different implementations might be equivalent for some starting states but differ for others.

Especially with growing research of quantum algorithms, leading to bigger and more complex circuits, testing and debugging quantum programs is becoming increasingly important.
The learning outcome of this step not only provides students with a deeper insight into an important tool for the verification of circuit transformations, but also emphasizes the importance of testing to new learners.

\begin{example}
    Revisiting the transformed circuit constructed in \autoref{ex:compilation}, we want to check its equivalence with the original circuit defined in~\autoref{fig:running-example}.
    After removing the non-unitary measurements from the transformed and original circuits, we construct the inverse of the original circuit by applying its gates in reverse order.
    The result of this process is shown in~\autoref{fig:verification}.
    We then store the measurement results in classical variables and simulate the verification circuit several times.
    As all of the measurement outcomes are equal to $0$, we can assume that the circuits are likely equal for the starting state $\ket{000}$.
    However, to check for the full equivalence of the circuits, we must further compare the outcomes for all potential starting states.
    We achieve this by iterating over all possible starting states in the computational basis and using the \block{Run quantum circuits starting from [STATE]} instruction.
    This evaluation shows that the circuits are not equivalent for some starting states, such as $\ket{010}$.
\end{example}

\subsection{Going Further}

To complete the education process for new learners interested in quantum design automation, they must also be provided with an access pass to the plethora of software tools currently available in the ecosystem.
These different tools provide a variety of functionalities, ranging from more sophisticated implementations of the strategies discussed in this section to error correction methods, HPC integration strategies and end user support software.
While the \emph{Scratch} programming platform provides an accessible playground that makes concepts easier to grasp for new learners, users should also be given the opportunity to move on to more complex tools as they gain experience with quantum computing.
To this end, as a final step, we show learners how the \block{Export to MLIR} block can be used to generate a textual representation of the implemented circuit.
This provides them with an interesting insight into the differences of the representations---including the verbosity of the MLIR-based IR.
It also gives learners an access path to the remaining ecosystem through further translations.

\begin{example}
    After exporting the constructed circuit, we store the generated MLIR representation locally.
    We then use it as input for the \emph{mqt-cc} compilation framework.
    Through its built-in QIR pipeline, we can then compile the program to a base profile QIR representation for a desired superconducting device topology.
    This representation can then be used to execute the circuit on the quantum hardware, successfully bridging the gap between the proposed \emph{Scratch}-based platform and the further quantum ecosystem down to actual physical execution.
\end{example}

\section{Evaluation}
\label{sec:evaluation}

To evaluate the pedagogical effectiveness of the proposed framework, we carried out a user study with a total of 24 users.
We asked them to complete a set of exercises and fill out a questionnaire to test the retained information.
The following first describes the methodology employed for the study, and then presents and discusses the obtained results.

\subsection{Methodology and Study Design}

We conducted a user study with post-graduate students from computer science and electrical engineering degrees.
These students had some fundamental understanding of quantum computing (such as the concepts of qubits and the effects of individual gates), but no extensive experience in quantum design automation.
To capture a realistic range of learning environments, the study was deployed in two distinct formats: an in-person, supervised laboratory session ($N = 13$) and an asynchronous, take-home \mbox{assignment ($N = 11$)}.

Both groups were provided with a set of exercises which required them to mirror the quantum design automation workflow outlined in Section~\ref{sec:exploring}---moving from initial circuit construction to resource estimation, optimization, and verification.
Upon completing the practical tasks, participants provided structured feedback through a questionnaire designed to assess their comprehension of the underlying concepts and their perceived usability of the block-based environment.
The questionnaire is divided into the following topic areas:

\paragraph*{Q1 -- Prior Experience} This group of questions evaluates the prior experience of participants in the field.
It asks participants to rate their expertise in quantum design automation topics such as resource estimation and verification, as well as their previous experience with the \emph{Scratch} programming platform on a Likert scale~\cite{likert1932}.
\paragraph*{Q2 -- Competence} Next, participants are tasked with answering a set of questions related to the concepts covered in the exercises.
This allows us to evaluate how well the exercises have conveyed these concepts to new~learners.
\paragraph*{Q3 -- Usability} Finally, this section asks the users for Likert-scale feedback on the ease of use of the proposed framework.
Participants are given the ability to rate their experience on a numerical scale and provide textual feedback.

\smallskip
Some participants also took part in an optional follow-up questionnaire one week later.
This follow-up once again asks the \emph{competence} questions to evaluate how well they have retained the conveyed concepts.

\subsection{User Study Results}

\begin{figure}
\centering
\begin{subfigure}{\columnwidth}
    \centering
    \input{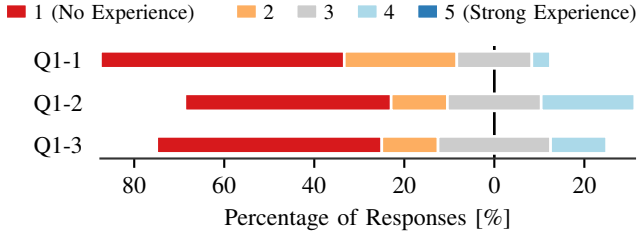}
    \caption{Self-reported previous experience of users with \emph{Scratch} (\emph{Q1-1}), circuit verification (\emph{Q1-2}), and resource estimation (\emph{Q3-1}).}
    \label{fig:eval-prev}
\end{subfigure}\\ \vspace{1.5em}

\begin{subfigure}{\columnwidth}
    \centering
    \input{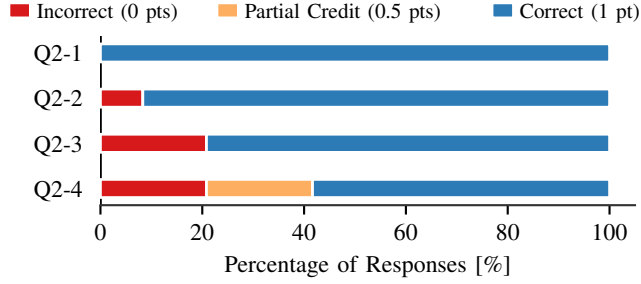}
    \caption{Assessment results after completing the exercises on reversing quantum circuits (\emph{Q2-1}), verification (\emph{Q2-2}), compilation (\emph{Q2-3}), and resource estimation (\emph{Q2-4}).}
    \label{fig:eval-test}

\end{subfigure}\\ \vspace{1.5em}

\begin{subfigure}{\columnwidth}
    \centering
    \input{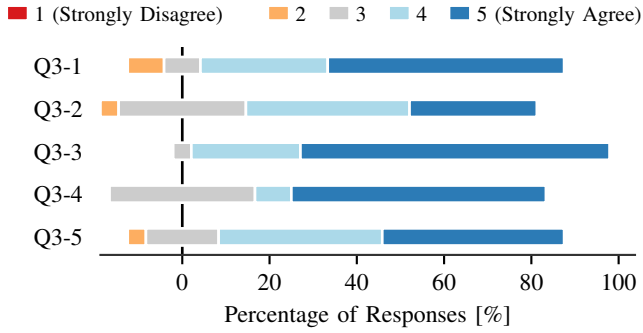}
    \caption{Self-reported user experience outcomes after conducting the study, indicating satisfaction with the block-based environment (\emph{Q3-1}), quantum-classical hybrid programs (\emph{Q3-2}), verification (\emph{Q3-3}), resource estimation (\emph{Q3-4}), and their confidence to explain the conveyed concepts to a fellow student (\emph{Q3-5}).}
    \label{fig:eval-ux}
\end{subfigure}
\caption{Results of the questionnaires filled out by the participants after completing the exercises.}
\end{figure}

To establish a baseline, participants self-reported their \emph{prior experience} with some of the covered concepts.
In particular, the questionnaire asks for the participants' experience with the \emph{Scratch} platform, quantum \emph{resource estimation} (RE), and quantum circuit \emph{verification}.
The corresponding results are illustrated in~\autoref{fig:eval-prev}.
They show that a majority of the participants did not have any major experience with either of the considered concepts.
In particular, more than 50\% of the participants did not report any experience with the \emph{Scratch} platform at all.
Similarly, a majority of participants only indicated limited prior experience with the quantum computing concepts covered by the exercises.
The median total reported prior expertise---obtained after averaging all individual question outcomes for each participant---was $1.83$ (close to ``\emph{little experience}'').
In general, the evaluation of this topic area shows that the participants previously only had limited contact with the topics covered in the exercises.

Conversely,~\autoref{fig:eval-test} shows the results of the \emph{competence} questions, gathered through assessment test questions as parts of the questionnaire.
\emph{Q2-1} to \emph{Q2-3} are single-choice questions that ask the participants to select one of four possible answers related to quantum circuit compilation and \emph{verification}.
Participants received full credits for selecting the correct answer and no points for selecting an incorrect answer.
\emph{Q2-4} is a multiple-choice question that requires participants to select potential use cases of \emph{resource estimation}.
Partial credits are given in cases, where participants did not select all correct choices but also did not select any incorrect choices.
The median total score of all participants was 3.5 out of 4 points, with a mean of $3.40$ and a standard deviation of $0.61$.
Furthermore, the median participant reported they were easily able to complete the exercise, with only some participants requiring significant additional time.
This shows that the exercises, carried out through the proposed framework, effectively conveyed the discussed topics, with most participants correctly responding to a majority of the questions. 

To gauge the proposed framework's \emph{usability}, we also asked them to self-report their experience during the completion of the tasks.
\autoref{fig:eval-ux} illustrates the outcomes of this topic.
In particular, a majority of participants was satisfied with the platform, either agreeing of strongly agreeing with its usefulness in learning quantum computing concepts.
Notably, for the process of inverting quantum circuits through the \emph{Scratch} extension, only one of the 24 participants selected ``\emph{neither agree nor disagree}'', while all other participants agreed or strongly agreed with their satisfaction.
More than half of the participants felt confident that they would be able to explain the covered concepts to their peers after the exercises.
In total, the median evaluation feedback for \emph{Q3} was $4.3$.
This indicates that the proposed framework provided a satisfactory user experience for the participants.

\bigskip

To evaluate the retention of the conveyed concepts, we performed a second round of \emph{competence} assessments one week after the exercises were performed.
A total of $N = 20$ participants took part of this second assessment.

\autoref{tab:retention} compares the assessment results obtained directly after the exercises with those in the retention test.
Notably, participants achieved equal or better results in all questions, except for \emph{Q2-3}, where the mean score fell by a total $4.2\%$.
As none of these differences are significant, with even a difference of $4.2\%$ only amounting to approximately one particpant, the test shows that participants were not only able to solve the test questions successfully directly after completing the exercises, but even after some time had already passed.

\medskip

As the study was carried out through different modalities---a supervised in-person session and a take-home exercise---we evaluated the framework's robustness across different learning environments by comparing the different cohorts.
The remote cohort achieved an overall mean \emph{competence} score of $3.36$ ($\text{SD} = 0.64$, $\text{MD} = 3.5$), closely mirroring the in-person mean score of $3.42$ ($\text{SD} = 0.61$, $\text{MD} = 3.5$).
Similarly, the remote cohort reported an overall median \emph{usability} score of $4.4$, while the in-person cohort reported a similar median score of $4.0$.

A \emph{Mann-Whitney} $U$ test~\cite{mann1947, mcknight2010} revealed no statistically significant difference in competence scores ($p = 0.831$) or perceived usability ($p = 0.134$) between the two groups. 
This strongly suggests that the proposed block-based framework is highly effective as a standalone, asynchronous educational tool, requiring no direct supervision to yield positive learning outcomes.

\medskip

Finally, the questionnaire also allowed participants to provide additional open-text feedback.
This feedback strongly supports the previous findings in this evaluation.
For instance, one participant notes: ``It is nice to have easily modifiable circuits and intuitive blocks provided by the quantum extension''.
Further participants specifically commented on performing equivalence checking through the exercises, mentioning, for instance: ``I liked the equivalence checking, because it visualized what it means to apply the reverse matrice very well''.
The questionnaire also asked for problems that arose during the completion of the tasks and usability friction points.
The main consensus of this question was that additional visualization methods would improve the clarity of the interface.
One user further asked to ``[...] provide pre-built blocks to print everything'', as the manual concatenation of measurement outcomes can be tedious for larger circuits.
Nonetheless, the participants largely agreed that the exercises were interesting and insightful, with one stating ``I liked working with Scratch and building simple circuits in an intuitive manner''.

\begin{table}
\caption{Concept Retention ($N = 20$)}
\label{tab:retention}
\begin{tabular}{l r r r}
Concept & Initial Pass Rate & 1-Week Pass Rate & $\Delta$ \\ \hline
Q2-1: Reversing Circuits & 100.0\% & 100.0\% & 0\% \\
Q2-2: Verification & 91.7\% & 95.0\% & 3.3\% \\
Q2-3: Compilation & 79.2\% & 75.0\% & -4.2\% \\
Q2-4: RE & 68.8\% & 70.0\% & 1.4\% \\
\end{tabular}
\vspace{1em}
\end{table}

\vspace{1em}
\subsection{Threats to Validity}

While the conducted evaluations have shown positive results, some additional factors must be considered when observing the outcomes of the questionnaires.
Firstly, the \emph{competence} questions consisted only of four single or multiple choice questions.
While this does not exhaustively capture the participants' full understanding of the concepts, the questions were selected specifically to cover the most important aspects of the exercises.
Furthermore, self-reported usability scores may be subject to bias.
However, the anonymity of questionnaire results mitigates this.
While students who participated in the study remotely may have utilized unauthorized external resources during the exercises and while filling out the questionnaire, the comparison of the two cohorts largely eliminates this concern, finding no significant statistical difference among them.
Additionally, because the retention test was optional, the 1-week follow-up results may be subject to self-selection bias, as more motivated students were more likely to complete it.

Finally, as all participants were post-graduate computer science and electrical engineering students, the obtained results may not generalize to more general learners.
Future work requires larger, cross-disciplinary cohorts to validate the proposed framework's usability across different levels of experience.

\section{Conclusion}
\label{sec:conclusion}

In this work, we presented a block-based framework for the development of quantum circuits as an educational tool for teaching quantum design automation concepts.
The framework is built as an extension of the \emph{Scratch} programming platform, allowing users to combine classical instructions for control flow, data storage, and user interfacing with quantum instructions for the creation of circuits.
Created circuits can be executed using a built-in simulator and analyzed by an integrated resource estimation procedure.
Through direct translations to other intermediate representations, created quantum circuits can further be used to facilitate the access to the wider quantum computing ecosystem.
A user study confirmed that, after the completion of a set of quantum design automation tasks, most participants gained a better understanding of quantum design automation concepts, with a majority of participants rating the framework's usability positively.
A web version of the proposed framework can be accessed online at~\link.

Future work includes the extension of the simulation backend to support a wider set of different simulator types, allowing learners to compare the performance of simulators for various quantum circuits, as well as the implementation of additional visualization methods within the framework to further increase accessibility.

	\vfill
\section*{Acknowledgments}
While preparing the manuscript, Claude Sonnet 4.6, Claude Opus 4.6, Gemini 3 Pro and gpt-5.3-codex were used through GitHub Copilot to improve readability, spelling, grammar, and clarity throughout the work. 
Each LLM output was reviewed by the authors and edited manually as needed. 
The authors take full responsibility for the final content.
The authors acknowledge funding from the European Research Council (ERC) under the European Union’s Horizon 2020 research and innovation program grant agreement No. 101001318,
as well as from the Munich Quantum Valley, which is supported by the Bavarian state government with funds from the Hightech Agenda Bayern Plus.
Furthermore, this work was supported by the BMFTR under grant No. 01MQ25001I (FullStaQD) and by BMIMI, BMWET, the State of Upper Austria and the State of Tyrol within the COMET module Quantum Algorithm Engineering (FFG grant No. 923923) managed by the Austrian Research Promotion Agency FFG.

\clearpage
\printbibliography

@STRING{tcad	= {{IEEE} Trans. on {CAD} of Integrated Circuits and Systems} }

@STRING{cacm	= {Comm. of the ACM} }

@STRING{sca_hpcasia = {Proceedings of the Supercomputing Asia and International Conference on High Performance Computing in Asia Pacific Region} }

@STRING{is	= {IEEE Software} }

@STRING{iccd	= {Int'l Conf. on Comp. Design} }

@STRING{iccad	= {Int'l Conf. on CAD} }

@STRING{date	= {Design, Automation and Test in Europe} }

@STRING{ispd	= {Int'l Symp. on {P}hysical {D}esign} }

@STRING{ismvl	= {Int'l Symp. on {M}ulti-{V}alued {L}ogic} }

@STRING{qsw = {{Int'l Conf. on Quantum Software}} }

@STRING{qce = {{Int'l Conf. on Quantum Computing and Engineering}} }

@STRING{qst   = {Quantum Science and Technology (QST)} }

@STRING{joss = {Journal of Open Source Software} }

@article{cross2022,
  title = {{{OpenQASM}}~3: {{A Broader}} and {{Deeper Quantum Assembly Language}}},
  author = {Cross, Andrew and {Javadi-Abhari}, Ali and Alexander, Thomas and De Beaudrap, Niel and Bishop, Lev S. and Heidel, Steven and Ryan, Colm A. and Sivarajah, Prasahnt and Smolin, John and Gambetta, Jay M. and Johnson, Blake R.},
  year = 2022,
  journal = {ACM Transactions on Quantum Computing}
}

@manual{qir,
  title         = {{QIR Specification}},
  author        = {{QIR Alliance}},
  year          = {2021},
  url           = {https://github.com/qir-alliance/qir-spec},
  note          = {see also \url{https://qir-alliance.org}}
}

@inproceedings{lattner2021,
  title = {{{MLIR}}: {{Scaling Compiler Infrastructure}} for {{Domain Specific Computation}}},
  booktitle = {{{IEEE}}/{{ACM International Symposium}} on {{Code Generation}} and {{Optimization}} ({{CGO}})},
  author = {Lattner, Chris and Amini, Mehdi and Bondhugula, Uday and Cohen, Albert and Davis, Andy and Pienaar, Jacques and Riddle, River and Shpeisman, Tatiana and Vasilache, Nicolas and Zinenko, Oleksandr},
  year = {2021},
}

@inproceedings{hopf2026,
	AUTHOR 		= {P. Hopf and E. Ochoa Lopez and Y. Stade and D. Rovara and N. Quetschlich and I. A. Florea and J. Izaac and R. Wille and L. Burgholzer},
	TITLE 		= {Integrating Quantum Software Tools with(in) {MLIR}},
	BOOKTITLE 	= sca_hpcasia,
	YEAR 		= {2026},
}

@inproceedings{burgholzer2026a,
	AUTHOR 		= {L. Burgholzer and D. Haag and Y. Stade and D. Rovara and Patrick Hopf and R. Wille},
	TITLE 		= {{The MQT Compiler Collection}: {A} Blueprint for a Future-Proof Quantum-Classical Compilation Framework},
	BOOKTITLE 	= date,
	YEAR 		= {2026},
}

@online{quirk,
  title = {{Quirk}},
  url = {https://github.com/Strilanc/Quirk},
  year = {2019},
  urldate = {2026-04-17}
}

@inproceedings{gallardo2024,
  title = {Quirk+: {{A Tool}} for {{Quantum Software Development Based}} on {{Quirk}}},
  booktitle = {2024 {{IEEE International Conference}} on {{Software Analysis}}, {{Evolution}} and {{Reengineering}} - {{Companion}} ({{SANER-C}})},
  author = {Gallardo, Javier Zayas and Moguel, Enrique and Canal, Carlos and {Garcia-Alonso}, Jose},
  year = 2024
}

@online{composer,
  title = {{IBM Quantum Composer}},
  url = {https://quantum.cloud.ibm.com/composer},
  year = {2016},
  urldate = {2026-04-17}
}

@incollection{gayathridevi2023,
  title = {Exploring {{IBM Quantum Experience}}},
  booktitle = {Quantum {{Computing}}: {{A Shift}} from {{Bits}} to {{Qubits}}},
  author = {Gayathri Devi, S. and Manjula Gandhi, S. and Chandia, S. and Boobalaragavan, P.},
  editor = {Pandey, Rajiv and Srivastava, Nidhi and Singh, Neeraj Kumar and Tyagi, Kanishka},
  year = 2023,
  publisher = {Springer Nature}
}

@article{corcoles2021,
  title = {Exploiting {{Dynamic Quantum Circuits}} in a {{Quantum Algorithm}} with {{Superconducting Qubits}}},
  author = {C{\'o}rcoles, A. D. and Takita, Maika and Inoue, Ken and Lekuch, Scott and Minev, Zlatko K. and Chow, Jerry M. and Gambetta, Jay M.},
  year = {2021},
  journal = {Physical Review Letters},
}

@article{pino2021,
  title = {Demonstration of the Trapped-Ion Quantum {{CCD}} Computer Architecture},
  author = {Pino, J. M. and Dreiling, J. M. and Figgatt, C. and Gaebler, J. P. and Moses, S. A. and Allman, M. S. and Baldwin, C. H. and {Foss-Feig}, M. and Hayes, D. and Mayer, K. and {Ryan-Anderson}, C. and Neyenhuis, B.},
  year = {2021},
  journal = {Nature},
}

@article{mcclure2016,
  title = {Rapid {{Driven Reset}} of a {{Qubit Readout Resonator}}},
  author = {McClure, D. T. and Paik, Hanhee and Bishop, L. S. and Steffen, M. and Chow, Jerry M. and Gambetta, Jay M.},
  year = {2016},
  journal = {Physical Review Applied},
}

@article{egger2018,
  title = {Pulsed {{Reset Protocol}} for {{Fixed-Frequency Superconducting Qubits}}},
  author = {Egger, D.J. and Werninghaus, M. and Ganzhorn, M. and Salis, G. and Fuhrer, A. and M{\"u}ller, P. and Filipp, S.},
  year = {2018},
  journal = {Physical Review Applied},
}

@misc{koch2025a,
  title = {{{HUGR}}: {{A Quantum-Classical Intermediate Representation}}},
  author = {Koch, Mark and Borgna, Agust{\'i}n and Sivarajah, Seyon and Lawrence, Alan and Edgington, Alec and Wilson, Douglas and Roy, Craig and Mondada, Luca and Heidemann, Lukas and Duncan, Ross},
  year = 2025,
  eprint = {2510.11420},
  primaryclass = {cs},
  archiveprefix = {arXiv}
}

@misc{koch2025,
  title = {{{GUPPY}}: {{Pythonic Quantum-Classical Programming}}},
  author = {Koch, Mark and Lawrence, Alan and Singhal, Kartik and Sivarajah, Seyon and Duncan, Ross},
  year = 2025,
  eprint = {2510.12582},
  primaryclass = {cs},
  archiveprefix = {arXiv}
}

@ARTICLE{ittah2024,
	AUTHOR    = {D. Ittah and A. Asadi and E. O. Lopez and S. Mironov and S. Banning and R. Moyard and M. J. Peng and J. Izaac},
	TITLE     = {Catalyst: a Python {JIT} compiler for auto-differentiable hybrid quantum programs},
	JOURNAL   = joss,
	YEAR      = {2024}
}

@article{ittah2022,
  title = {{{QIRO}}: {{A Static Single Assignment-based Quantum Program Representation}} for {{Optimization}}},
  author = {Ittah, David and H{\"a}ner, Thomas and Kliuchnikov, Vadym and Hoefler, Torsten},
  year = 2022,
  journal = {ACM Transactions on Quantum Computing}
}

@misc{kirin,
  title         = {{Kirin}: Kernel Intermediate Representation Infrastructure},
  author        = {{QuEra Computing Inc}},
  year          = {2025},
  url           = {https://queracomputing.github.io/kirin/latest/},
}

@inproceedings{grover,
    author = {Lov K. Grover},
    booktitle = {Theory of computing},
    pages = {212--219},
    title = {A fast quantum mechanical algorithm for database search},
    year = {1996}
}

@online{jeff,
  author = {Jeff contributors},
  title = {{Jeff Bridges Compilation}},
  year = 2025,
  url = {https://github.com/unitaryfoundation/jeff},
  urldate = {2026-04-13}
}

@misc{seidel2024,
  title = {Qrisp: {{A Framework}} for {{Compilable High-Level Programming}} of {{Gate-Based Quantum Computers}}},
  author = {Seidel, Raphael and Bock, Sebastian and Zander, Ren{\'e} and Petri{\v c}, Matic and Steinmann, Niklas and Tcholtchev, Nikolay and Hauswirth, Manfred},
  year = {2024},
  archivePrefix={arXiv},
  eprint = {2406.14792},
  primaryclass = {quant-ph},
}

@article{meuli2020,
  title = {Enabling Accuracy-Aware {{Quantum}} Compilers Using Symbolic Resource Estimation},
  author = {Meuli, Giulia and Soeken, Mathias and Roetteler, Martin and H{\"a}ner, Thomas},
  year = 2020,
  journal = {Proc. ACM Program. Lang.}
}

@misc{qiskit,
    author = {{Qiskit contributors}},
    title = {Qiskit: An Open-source Framework for Quantum Computing},
    year = {2023},
    doi = {10.5281/zenodo.2573505}
}

@misc{ransford2025,
  title = {{Helios}: A 98-qubit trapped-ion quantum computer},
  author = {Anthony Ransford and M. S. Allman and Jake Arkinstall and J. P. Campora III and Samuel F. Cooper et al.},
  year = 2025,
  eprint = {2511.05465},
  primaryclass = {quant-ph},
  archiveprefix = {arXiv}
}

@book{nielsen2010,
  title     = {Quantum Computation and Quantum Information},
  author    = {Michael A. Nielsen and Isaac L. Chuang},
  year      = {2010},
  publisher = {Cambridge University Press},
}

@article{tket,
	year = 2020,
	author = {Seyon Sivarajah and Silas Dilkes and Alexander Cowtan and Will Simmons and Alec Edgington and Ross Duncan},
	title = {t$\vert$ket{\rangle}: a retargetable compiler for {NISQ} devices},
	journal = {Quantum Science and Technology},
}

@article{burgholzer2021a,
  title = {Advanced {{Equivalence Checking}} for {{Quantum Circuits}}},
  author = {Burgholzer, Lukas and Wille, Robert},
  year = 2021,
  journal = {IEEE Transactions on Computer-Aided Design of Integrated Circuits and Systems}
}

@inproceedings{yamashita2010,
  title = {Fast Equivalence-Checking for Quantum Circuits},
  booktitle = {2010 {{IEEE}}/{{ACM International Symposium}} on {{Nanoscale Architectures}}},
  author = {Yamashita, Shigeru and Markov, Igor L.},
  year = 2010
}

@article{resnick2009,
  title = {Scratch: Programming for All},
  author = {Resnick, Mitchel and Maloney, John and {Monroy-Hern{\'a}ndez}, Andr{\'e}s and Rusk, Natalie and Eastmond, Evelyn and Brennan, Karen and Millner, Amon and Rosenbaum, Eric and Silver, Jay and Silverman, Brian and Kafai, Yasmin},
  year = 2009,
  journal = cacm
}

@article{maloney2010,
  title = {The {{Scratch Programming Language}} and {{Environment}}},
  author = {Maloney, John and Resnick, Mitchel and Rusk, Natalie and Silverman, Brian and Eastmond, Evelyn},
  year = 2010,
  journal = {ACM Trans. Comput. Educ.}
}

@article{armoni2015,
  title = {From {{Scratch}} to ``{{Real}}'' {{Programming}}},
  author = {Armoni, Michal and {Meerbaum-Salant}, Orni and {Ben-Ari}, Mordechai},
  year = 2015,
  journal = {ACM Trans. Comput. Educ.}
}

@misc{bergholm2022,
  title={PennyLane: Automatic differentiation of hybrid quantum-classical computations}, 
  author={Ville Bergholm and Josh Izaac and Maria Schuld and Christian Gogolin and Shahnawaz Ahmed and Vishnu Ajith and M. Sohaib Alam and Guillermo Alonso-Linaje and B. AkashNarayanan and Ali Asadi and Juan Miguel Arrazola and Utkarsh Azad and Sam Banning and Carsten Blank and Thomas R Bromley and Benjamin A. Cordier and Jack Ceroni and Alain Delgado and Olivia Di Matteo and Amintor Dusko and Tanya Garg and Diego Guala and Anthony Hayes and Ryan Hill and Aroosa Ijaz and Theodor Isacsson and David Ittah and Soran Jahangiri and Prateek Jain and Edward Jiang and Ankit Khandelwal and Korbinian Kottmann and Robert A. Lang and Christina Lee and Thomas Loke and Angus Lowe and Keri McKiernan and Johannes Jakob Meyer and J. A. Montañez-Barrera and Romain Moyard and Zeyue Niu and Lee James O'Riordan and Steven Oud and Ashish Panigrahi and Chae-Yeun Park and Daniel Polatajko and Nicolás Quesada and Chase Roberts and Nahum Sá and Isidor Schoch and Borun Shi and Shuli Shu and Sukin Sim and Arshpreet Singh and Ingrid Strandberg and Jay Soni and Antal Száva and Slimane Thabet and Rodrigo A. Vargas-Hernández and Trevor Vincent and Nicola Vitucci and Maurice Weber and David Wierichs and Roeland Wiersema and Moritz Willmann and Vincent Wong and Shaoming Zhang and Nathan Killoran},
  year={2022},
  archivePrefix={arXiv},
  eprint={1811.04968},
  primaryClass={quant-ph},
}

@misc{vandam2023using,
  title={Using Azure Quantum Resource Estimator for Assessing Performance of Fault Tolerant Quantum Computation}, 
  author={Wim van Dam and Mariia Mykhailova and Mathias Soeken},
  year={2023},
  eprint={2311.05801},
  archivePrefix={arXiv},
  primaryclass = {quant-ph},
}

@incollection{prateek2023,
  title = {Quantum {{Programming}} on~{{Azure Quantum}}---{{An Open Source Tool}} for~{{Quantum Developers}}},
  booktitle = {Quantum {{Computing}}: {{A Shift}} from {{Bits}} to {{Qubits}}},
  author = {Prateek, Kumar and Maity, Soumyadev},
  editor = {Pandey, Rajiv and Srivastava, Nidhi and Singh, Neeraj Kumar and Tyagi, Kanishka},
  year = 2023,
  publisher = {Springer Nature}
}

@inproceedings{upadhyay2025a,
  title = {Analyzing the {{Evolution}} and {{Maintenance}} of {{Quantum Software Repositories}}},
  booktitle = {2025 {{IEEE International Conference}} on {{Quantum Software}} ({{QSW}})},
  author = {Upadhyay, Krishna and Chhetri, Vinaik and Siddique, A.B. and Farooq, Umar},
  year = 2025
}

@inproceedings{dimatteo2024,
	title = {On the need for effective tools for debugging quantum programs},
	booktitle = {International Workshop on Quantum Software Engineering},
	author = {Di Matteo, Olivia},
  year = {2024},
}

@article{metwalli2024,
  title = {Testing and {{Debugging Quantum Circuits}}},
  author = {Metwalli, Sara Ayman and Van Meter, Rodney},
  year = {2024},
  journal = {IEEE Transactions on Quantum Engineering},
}

@inproceedings{huang2019,
	title = {Statistical assertions for validating patterns and finding bugs in quantum programs},
	booktitle = {International Symposium on Computer Architecture},
	author = {Huang, Yipeng and Martonosi, Margaret},
  year = {2019},
}

@misc{miranskyy2021,
	title = {On Testing and Debugging Quantum Software},
	author = {Miranskyy, Andriy and Zhang, Lei and Doliskani, Javad},
  year = {2021},
  archivePrefix={arXiv},
  primaryclass = {quant-ph},
	eprint = {2103.09172},
}

@article{garciadelabarrera2023,
	title = {Quantum software testing: State of the art},
	journaltitle = {Journal of Software: Evolution and Process},
	author = {García de la Barrera, Antonio and García-Rodríguez de Guzmán, Ignacio and Polo, Macario and Piattini, Mario},
  year = {2023},
}

@misc{li2020,
  title = {Proq: {{Projection-based Runtime Assertions}} for {{Debugging}} on a {{Quantum Computer}}},
  author = {Li, Gushu and Zhou, Li and Yu, Nengkun and Ding, Yufei and Ying, Mingsheng and Xie, Yuan},
  year = {2020},
  archivePrefix={arXiv},
  eprint = {1911.12855},
  primaryclass = {quant-ph},
}

@misc{ying2022,
	title = {Birkhoff-von {N}eumann Quantum Logic as an Assertion Language for Quantum Programs},
	number = {{arXiv}:2205.01959},
	author = {Ying, Mingsheng},
	date = {2022-05-04},
	eprint = {2205.01959 [quant-ph]},
  archivePrefix={arXiv},
  primaryclass = {quant-ph},
}

@inproceedings{liu2020,
	title = {Quantum Circuits for Dynamic Runtime Assertions in Quantum Computation},
	booktitle = {International Conference on Architectural Support for Programming Languages and Operating Systems},
	author = {Liu, Ji and Byrd, Gregory T. and Zhou, Huiyang},
  year = {2020},
}

@inproceedings{liu2021,
	title = {Systematic Approaches for Precise and Approximate Quantum State Runtime Assertion},
	booktitle = {International Symposium on High-Performance Computer Architecture ({HPCA})},
	author = {Liu, Ji and Zhou, Huiyang},
  year = {2021},
}

@misc{witharana2023,
	title = {{quAssert}: Automatic Generation of Quantum Assertions},
  number = {{arXiv}:2303.01487},
	publisher = {{arXiv}},
	author = {Witharana, Hasini and Volya, Daniel and Mishra, Prabhat},
	date = {2023-03-02},
  archivePrefix={arXiv},
  primaryclass = {quant-ph},
	eprint = {2303.01487},
}

@inproceedings{rovara2025a,
	AUTHOR    = {D. Rovara and L. Burgholzer and R. Wille},
	TITLE     = {{A Framework for the Efficient Evaluation of Runtime Assertions on Quantum Computers}},
	BOOKTITLE = qsw,
	YEAR      = {2026},
}

@INPROCEEDINGS{rovara2025b,
	AUTHOR    = {D. Rovara and L. Burgholzer and R. Wille},
	TITLE     = {{Automatically Refining Assertions for Efficient Debugging of Quantum Programs}},
	BOOKTITLE = qce,
	YEAR      = {2025},
}

@INPROCEEDINGS{rovara2025c,
	AUTHOR    = {D. Rovara and L. Burgholzer and R. Wille},
	TITLE     = {{A Framework for Debugging Quantum Programs}},
	BOOKTITLE = qsw,
	YEAR      = {2025},
}

@inproceedings{zhou2019,
  title = {An applied quantum {{Hoare}} logic},
  booktitle = {Conference on {{Programming Language Design}} and {{Implementation}}},
  author = {Zhou, Li and Yu, Nengkun and Ying, Mingsheng},
  date = {2019-06-08},
}

@INPROCEEDINGS{forster2025a,
    AUTHOR    = {T. Forster and N. Quetschlich and M. Soeken and R. Wille},
    TITLE     = {{Improving Hardware Requirements for Fault-Tolerant Quantum Computing by Optimizing Error Budget Distributions}},
    BOOKTITLE = qce,
    YEAR      = {2025},
}

@INPROCEEDINGS{forster2025b,
    AUTHOR    = {T. Forster and N. Quetschlich and R. Wille},
    TITLE     = {{Quantum Circuit Optimization for the Fault-Tolerance Era: Do We Have to Start from Scratch?}},
    BOOKTITLE = qce,
    YEAR      = {2025},
}

@article{mann1947,
  title = {On a {{Test}} of {{Whether}} One of {{Two Random Variables}} Is {{Stochastically Larger}} than the {{Other}}},
  author = {Mann, H. B. and Whitney, D. R.},
  year = 1947,
  journal = {The Annals of Mathematical Statistics},
  publisher = {Institute of Mathematical Statistics}
}

@incollection{mcknight2010,
  title = {Mann-{{Whitney U Test}}},
  booktitle = {The {{Corsini Encyclopedia}} of {{Psychology}}},
  author = {McKnight, Patrick E. and Najab, Julius},
  year = 2010,
  publisher = {John Wiley \& Sons, Ltd}
}

@article{likert1932,
  title = {A Technique for the Measurement of Attitudes},
  author = {Likert, R.},
  year = 1932,
  journal = {Archives of Psychology}
}

@inproceedings{willeMQTHandbookSummary2024,
  title = {The {{MQT Handbook}}: {{A Summary}} of {{Design Automation Tools}} and {{Software}} for {{Quantum Computing}}},
  booktitle = {{IEEE} International Conference on Quantum Software ({QSW})},
  author = {Wille, Robert and Berent, Lucas and Forster, Tobias and Kunasaikaran, Jagatheesan and Mato, Kevin and Peham, Tom and Quetschlich, Nils and Rovara, Damian and Sander, Aaron and Schmid, Ludwig and Schoenberger, Daniel and Stade, Yannick and Burgholzer, Lukas},
  date = {2024},
}

@inproceedings{brand1993,
  title = {Verification of large synthesized designs},
  booktitle = iccad,
  author = {Brand, Daniel},
  date = {1993},
  pages = {534--537},
  eventtitle = {iccad}
}

@book{molitor2010,
  title = {Equivalence checking of digital circuits: {{Fundamentals}}, principles, methods},
  author = {Molitor, Paul and Mohnke, Janett},
  date = {2010},
  publisher = {{Springer}}
}

@inproceedings{marques1999,
  title = {Combinational equivalence checking using satisfiability and recursive learning},
  booktitle = date,
  author = {Marques-Silva, João and Glass, Thomas},
  date = {1999},
}

@inproceedings{jha1997,
  title = {Equivalence checking using abstract {{BDDs}}},
  booktitle = iccd,
  author = {Jha, S. and Lu, Y. and Minea, M. and Clarke, E. M.},
  date = {1997},
  eventtitle = {iccd}
}

@book{wang2009,
  title = {Electronic {{Design Automation}}: {{Synthesis}}, {{Verification}}, and {{Test}}},
  author = {Wang, Laung-Terng and Chang, Yao-Wen and Cheng, Kwang-Ting (Tim)},
  year = 2009,
  publisher = {Morgan Kaufmann},
  googlebooks = {3XBe7dLb5NEC}
}

@article{li2023,
  title = {{{QASMBench}}: {{A Low-Level Quantum Benchmark Suite}} for {{NISQ Evaluation}} and {{Simulation}}},
  author = {Li, Ang and Stein, Samuel and Krishnamoorthy, Sriram and Ang, James},
  year = 2023,
  journal = {ACM Transactions on Quantum Computing}
}

@article{quetschlich2023,
  title = {{{MQT Bench}}: {{Benchmarking Software}} and {{Design Automation Tools}} for {{Quantum Computing}}},
  author = {Quetschlich, Nils and Burgholzer, Lukas and Wille, Robert},
  year = 2023,
  journal = {Quantum},
  publisher = {Verein zur F\"orderung des Open Access Publizierens in den Quantenwissenschaften}
}

@article{quetschlich2025,
  title = {{{MQT Predictor}}: {{Automatic Device Selection}} with {{Device-Specific Circuit Compilation}} for {{Quantum Computing}}},
  author = {Quetschlich, Nils and Burgholzer, Lukas and Wille, Robert},
  year = 2025,
  journal = {ACM Transactions on Quantum Computing}
}

@article{alexeev2025,
  title = {Artificial Intelligence for Quantum Computing},
  author = {Alexeev, Yuri and Farag, Marwa H. and Patti, Taylor L. and Wolf, Mark E. and Ares, Natalia and {Aspuru-Guzik}, Al{\'a}n and Benjamin, Simon C. and Cai, Zhenyu and Cao, Shuxiang and Chamberland, Christopher and Chandani, Zohim and Fedele, Federico and Hamamura, Ikko and Harrigan, Nicholas and Kim, Jin-Sung and Kyoseva, Elica and Lietz, Justin G. and Lubowe, Tom and McCaskey, Alexander and Melko, Roger G. and Nakaji, Kouhei and Peruzzo, Alberto and Rao, Pooja and Schmitt, Bruno and Stanwyck, Sam and Tubman, Norm M. and Wang, Hanrui and Costa, Timothy},
  year = 2025,
  journal = {Nature Communications},
  publisher = {Nature Publishing Group}
}

@misc{fosel2021a,
  title = {Quantum Circuit Optimization with Deep Reinforcement Learning},
  author = {F{\"o}sel, Thomas and Niu, Murphy Yuezhen and Marquardt, Florian and Li, Li},
  year = 2021,
  eprint = {2103.07585},
  primaryclass = {quant-ph},
  archiveprefix = {arXiv}
}

@article{moro2021,
  title = {Quantum Compiling by Deep Reinforcement Learning},
  author = {Moro, Lorenzo and Paris, Matteo G. A. and Restelli, Marcello and Prati, Enrico},
  year = 2021,
  journal = {Communications Physics},
  publisher = {Nature Publishing Group}
}

@article{escanez2025,
  title = {{{QScratch}}: Introduction to Quantum Mechanics Concepts through Block-Based Programming},
  author = {{Escanez-Exposito}, Daniel and {Rodriguez-Vega}, Marcos and {Rosa-Remedios}, Carlos and {Caballero-Gil}, Pino},
  year = 2025,
  journal = {EPJ Quantum Technology}
}

@inproceedings{rosa2024,
  title = {{{WIP}}: {{Teaching Basic Concepts}} of {{Quantum Computing Using Scratch}}},
  booktitle = {2024 {{IEEE Frontiers}} in {{Education Conference}} ({{FIE}})},
  author = {{Rosa-Remedios}, Carlos and {Caballero-Gil}, Pino and {Escanez-Exposito}, Daniel},
  year = 2024
}

@inproceedings{suchara2013,
  title = {{{QuRE}}: {{The Quantum Resource Estimator}} Toolbox},
  booktitle = {{{IEEE}} {{International Conference}} on {{Computer Design}} ({{ICCD}})},
  author = {Suchara, Martin and Kubiatowicz, John and Faruque, Arvin and Chong, Frederic T. and Lai, Ching-Yi and Paz, Gerardo},
  year = 2013
}

@INPROCEEDINGS{rovara2026,
    AUTHOR    = {D. Rovara and L. Burgholzer and R. Wille},
    TITLE     = {{Quantum Hardware-Efficient Selection of Auxiliary Variables for QUBO Formulations}},
    BOOKTITLE = date,
    YEAR      = {2026},
}

@ARTICLE{schmid2024,
	AUTHOR    = {L. Schmid and D. Locher and M. Rispler and S. Blatt and J. Zeiher and M. Müller and R. Wille},
	TITLE     = {{Computational Capabilities and Compiler Development for Neutral Atom Quantum processors—Connecting Tool Developers and Hardware Experts}},
	JOURNAL   = qst,
	YEAR      = 2024,
}

@INPROCEEDINGS{li2019a,
  title = {Tackling the {{Qubit Mapping Problem}} for {{NISQ-Era Quantum Devices}}},
  booktitle = {{{International Conference}} on {{Architectural Support}} for {{Programming Languages}} and {{Operating Systems}}},
  author = {Li, Gushu and {View Profile} and Ding, Yufei and {View Profile} and Xie, Yuan and {View Profile}},
  year = 2019
}

@inproceedings{zilk2022,
  title = {A Compiler for Universal Photonic Quantum Computers},
  booktitle = {{{IEEE}}/{{ACM International Workshop}} on {{Quantum Computing Software}} ({{QCS}})},
  author = {Zilk, Felix and Staudacher, Korbinian and Guggemos, Tobias and F{\"u}rlinger, Karl and Kranzlm{\"u}ller, Dieter and Walther, Philip},
  year = 2022
}

@INPROCEEDINGS{rovara2026a,
  AUTHOR    = {D. Rovara and N. Quetschlich and R. Wille},
	TITLE     = {{A Framework to Formulate Pathfinding Problems for Quantum Computing}},
	BOOKTITLE = ismvl,
	YEAR      = {2026},
}

@INPROCEEDINGS{quetschlich2024,
	AUTHOR    = {N. Quetschlich and T. Forster and A. Osterwind and D. Helms and R. Wille},
	TITLE     = {{Towards Equivalence Checking of Classical Circuits Using Quantum Computing}},
	BOOKTITLE = qce,
	YEAR      = {2024},
}

@INPROCEEDINGS{wille2023,
    AUTHOR    = {R. Wille and L. Burgholzer},
    TITLE     = {{MQT QMAP: Efficient Quantum Circuit Mapping}},
    BOOKTITLE = ispd,
    YEAR      = {2023},
}

@misc{rovara2025d,
	AUTHOR    = {D. Rovara and L. Burgholzer and R. Wille},
	TITLE     = {{Qubit Reuse Beyond Reorder and Reset: Optimizing Quantum Circuits by Fully Utilizing the Potential of Dynamic Circuits}},
	YEAR      = {2025},
  archivePrefix={arXiv},
	EPRINT    = {2511.22712},
  primaryclass = {quant-ph},
}

@inproceedings{tudisco2026,
	AUTHOR    = {A. Tudisco and P. Hopf and L Schulte and D. Volpe and G. Turvani and R. Wille},
	TITLE     = {{Fidelity-Based Quantum Device Selection Using Graph Neural Networks}},
	BOOKTITLE = qsw,
	YEAR      = {2026},
}

@misc{kang2025,
  title = {{{PennyLane-Lightning MPI}}: {{A}} Massively Scalable Quantum Circuit Simulator Based on Distributed Computing in {{CPU}} Clusters},
  author = {Kang, Ji-Hoon and Ryu, Hoon},
  year = 2025,
  eprint = {2508.13615},
  primaryclass = {quant-ph},
  publisher = {arXiv},
  archiveprefix = {arXiv}
}

@inproceedings{bayraktar2023,
  title = {{{cuQuantum SDK}}: {{A High-Performance Library}} for {{Accelerating Quantum Science}}},
  booktitle = qce,
  author = {Bayraktar, Harun and Charara, Ali and Clark, David and Cohen, Saul and Costa, Timothy and Fang, Yao-Lung L. and Gao, Yang and Guan, Jack and Gunnels, John and Haidar, Azzam and Hehn, Andreas and Hohnerbach, Markus and Jones, Matthew and Lubowe, Tom and Lyakh, Dmitry and Morino, Shinya and Springer, Paul and Stanwyck, Sam and Terentyev, Igor and Varadhan, Satya and Wong, Jonathan and Yamaguchi, Takuma},
  year = 2023
}

@misc{javadi-abhari2024,
  title = {Quantum Computing with {{Qiskit}}},
  author = {{Javadi-Abhari}, Ali and Treinish, Matthew and Krsulich, Kevin and Wood, Christopher J. and Lishman, Jake and Gacon, Julien and Martiel, Simon and Nation, Paul D. and Bishop, Lev S. and Cross, Andrew W. and Johnson, Blake R. and Gambetta, Jay M.},
  year = 2024,
  eprint = {2405.08810},
  primaryclass = {quant-ph},
  publisher = {arXiv},
  archiveprefix = {arXiv}
}

@ARTICLE{zulehner2018,
	AUTHOR    = {A. Zulehner and R. Wille},
	TITLE     = {{Advanced Simulation of Quantum Computations}},
	JOURNAL   = tcad,
	YEAR      = 2018,
}

@ARTICLE{grurl2022,
	AUTHOR    = {T. Grurl and J. Fu{\ss} and R. Wille},
	TITLE     = {{Noise-aware Quantum Circuit Simulation With Decision Diagrams}},
	JOURNAL   = tcad,
	YEAR      = 2022,
}

\end{document}